\documentclass[11pt]{article}
\usepackage[a4paper, portrait, margin=2.5cm]{geometry}
\usepackage{graphicx}
\usepackage{lastpage}
\usepackage{fancyhdr}
\usepackage{siunitx} 
\usepackage{hyperref} 
\usepackage{amssymb,amsmath}

\fancypagestyle{firstpage}{%
  \setlength\headheight{154pt}
  \fancyhf{}%
  \fancyhead[C]{\includegraphics[width=\textwidth]{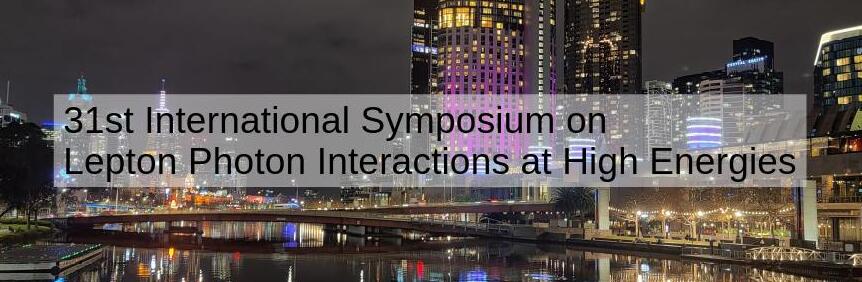}}%
  \fancyfoot[C]{CC-BY-4.0 licence}%
}

\begin{document}
\title{Latest results from Daya Bay using the full dataset} 
\date{17-21 July 2023}
\author{Zhiyuan Chen \\ Institute of High Energy Physics, CAS, Beijing 100049, China \\ \textit{On behalf of the Daya Bay collaboration}} 

\newgeometry{top=2cm, bottom=7cm}
\maketitle
\thispagestyle{firstpage}
\abstract{
The Daya Bay Reactor Neutrino Experiment was designed with the primary goal of precisely measuring the neutrino mixing parameter, $\theta_{13}$. Eight identically-designed gadolinium-doped liquid scintillator detectors installed in three underground experimental halls measure the reactor antineutrinos from six nuclear reactors at different distances. Until its shutdown at the end of 2020, Daya Bay experiment has acquired nearly 6 million inverse beta decay candidates with neutron captured on gadolinium. In this talk, the latest neutrino oscillation analysis results based on full data will be presented. The resulting oscillation parameters are $\sin^{2}2\theta_{13} = 0.0851 \pm 0.0024$, $\Delta m^{2}_{32} = (2.466 \pm 0.060) \times 10^{-3} \si{eV}^{2}$ for the normal mass ordering or $\Delta m^{2}_{32} = -(2.571 \pm 0.060) \times 10^{-3} \si{eV}^{2}$ for the inverted mass ordering, which are the most precise measurement of $\theta_{13}$ and $\Delta m^{2}_{32}$ so far. Moreover, latest results on other topics such as the search of high energy reactor neutrino is included as well.
}

\restoregeometry
\section{The Daya Bay Reactor Neutrino Experiment}
The Daya Bay Reactor Neutrino Experiment was designed to measure the mixing angle $\theta_{13}$  via the investigation of reactor antineutrino disappearance at a about 2 km baseline resulting from neutrino oscillation. 
It began data taking in late 2011 and finished operation at the end of 2020. 

\begin{figure}
    \centering
    \includegraphics[height=0.4\linewidth]{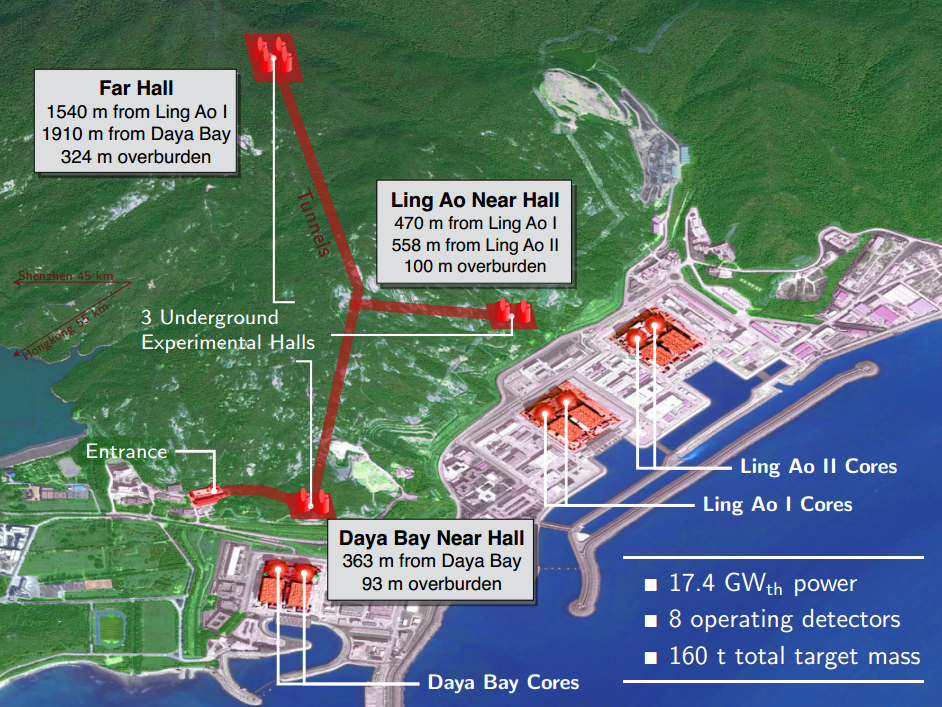}
    \includegraphics[height=0.4\linewidth]{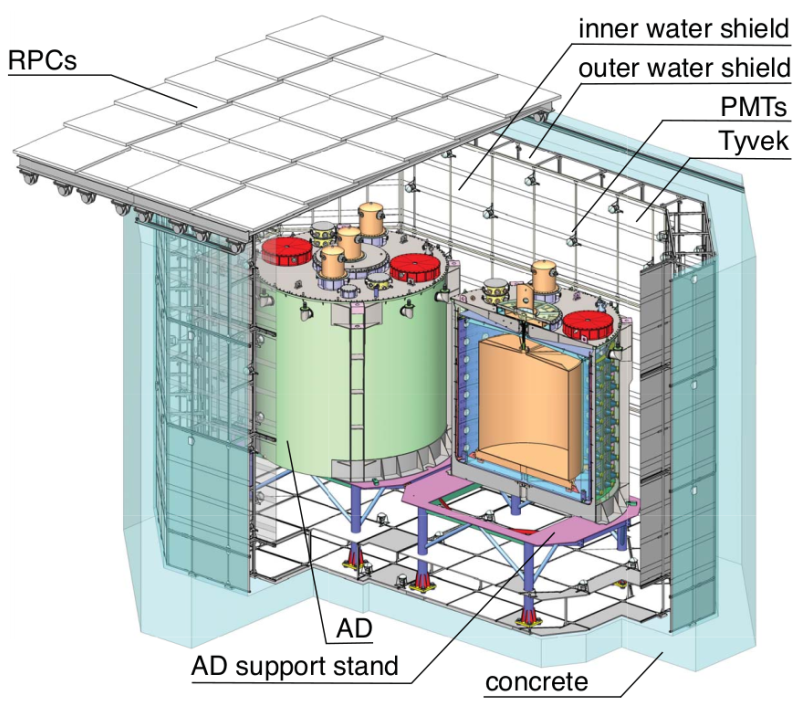}
    \caption{(Left) Layout of the Daya Bay experiment. Two near experimental halls, EH1 and EH2, monitor reactor neutrino flux and spectrum, while the far hall EH3 observes the oscillation driven by $\theta_{13}$ mixing angle. (Right) Sketch of the detectors in one of the near halls. The ADs are installed in a water pool and covered with RPCs.}
    \label{fig:layout}
\end{figure}

To reduce systematic issues, Daya Bay performed relative measurement with Far/Near ratio. 
As shown in Figure \ref{fig:layout} (left), Daya Bay used eight antineutrino detectors (ADs) to detect $\overline{\nu}_{e}$s emitted from the reactors at the Daya Bay-Ling Ao nuclear power facility in China. 
The ADs were installed in three underground experimental halls and were submerged in water pools to reduce ambient radiation, shown in Fig. \ref{fig:layout} (right). 
Each pool was optically divided into inner (IWS) and outer (OWS) water Cherenkov detectors for detecting cosmic-ray muons as muon veto systems. 
Four layers of resistive plate chambers (RPCs) cover the top of each water pool to provide another independent muon detector. 
As an neutrino target, 20 tonnes of liquid scintillator doped with 0.1\% gadolinium by weight (GdLS) was contained in a 3-m-diameter acrylic cylinder enclosed inside a 4-m-diameter acrylic cylinder filled with 22 tonnes of undoped liquid scintillator (LS) in each AD. 
192 photomultiplier tubes (PMTs) were installed on the barrel surface of the AD to detect optical photons generated in the scintillator.

Reactor antineutrinos are detected via the inverse beta decay (IBD) reaction: $\overline{\nu}_{e} + p \to e^{+} + n$. Positron deposits the energy quickly and forms the prompt signal. Neutron is captured on a nucleus and becomes the delayed signal. With the coincidence of prompt and delayed signals, we are able to suppress the backgrounds remarkably.

\section{Neutrino Oscillation Results}
In this section, we report a new measurement of $\sin^{2}2\theta_{13}$ and $\Delta m^{2}_{32}$ using a final sample of $5.55 \times 10^6$ IBD candidates with the final-state neutron captured on gadolinium (nGd) acquired in 3158 days of operation \cite{nGd2023}.
Details of the analysis process and techniques can be found in Refs. \cite{nGd2017PRD,nGd2018}. In this Proceeding, we focus on the improvements to the analysis techniques.

In terms of the energy response of the detector, a correction for the nonlinear response of the electronics was applied to each channel. This correction was derived from the waveform output from a flash-ADC readout system running in parallel with the default ADC system of EH1-AD1 in 2016 \cite{FADC2018}.

A new source of PMT flashers was observed in the 7-AD operation period that were not rejected by the previous criteria. By utilizing the characteristic charge pattern and temporal distribution of these new flashers, we proposed additional selection criteria that removed over 99\% of this instrumental background with a negligible IBD selection inefficiency.


The largest correlated background is $\beta$-n decay of cosmogenic radioisotopes $^{9}$Li/$^{8}$He. In order to determine this background, we paired muons with all IBD candidates within $\pm$2 s. To improve discrimination of $^{9}$Li/$^{8}$He from other processes, candidate events were separated into several samples based on the visible energy deposited by the muon in the AD and the distance between the prompt and delayed signals, $\Delta r$. The rates and energy spectra of the dominant cosmogenic radioisotopes were extracted with a simultaneous fit to 12 two-dimensional histograms defined by the different muon samples in the three experimental halls for the two $\Delta r$ regions. We simply measured the sum of these two radioisotopes, in consideration of the comparable lifetimes of $^{9}$Li and $^{8}$He. This method provides higher statistics and a better determination of the low-energy part of the $\beta$ spectrum of $^{9}$Li/$^{8}$He than the previous determination while reducing the rate uncertainty to less than 25\%.

Due to the gradual loss of functional PMTs near the top of the water pools, the muon detection efficiency of water pools dropped with time, particularly in the 7-AD period. As a consequence, a new muon-induced background, named as “muon-x” became significant.
The muon-x background was caused by low energy muons that passed through the IWS undetected. 
These events typically consisted of the muon as the prompt signal and a Michel electron from muon decay a product of muon capture or a spallation neutron as the delayed signal. 
Muon-x background was efficiently suppressed by rejecting events with a delayed signal less than 410 $\mu$s after a muon identified with a more stringent IWS PMT-hit multiplicity requirement of 6 \textless nHit $\le$ 12 which led to a \textless 0.1\% loss in livetime. 
To determine the rate of these two backgrounds, the prompt-energy spectra of the IBD-candidate sample were extended to 250 MeV and were fitted to the spectra of the previously described fast-neutron sample and the muon-x sample with IWS nHit = 7. 
Through extrapolation, we obtained their rates in the range of 0.7 MeV \textless $E_p$ \textless 12 MeV.

We extracted the oscillation parameters using the survival probability of three-flavor oscillation given by
\begin{equation}
\begin{aligned}
    P = & 1-\cos^{4}\theta_{13}\sin^{2}2\theta_{12}\sin^{2}\Delta_{21} \\
    & -\sin^{2}2\theta_{13}(\cos^{2}\theta_{12}\sin^{2}\Delta_{31} + \sin^{2}\theta_{12}\sin^{2}\Delta_{32}),
\end{aligned}
\end{equation}
where $\Delta_{ij}=\Delta m^{2}_{ij}L/(4\hbar cE)$ with $\Delta m^{2}_{ij}$ in eV$^2$, \textit{L} is the baseline in meters between an AD and a reactor core and \textit{E} is the energy of the $\overline{\nu}_{e}$ in MeV. Alternatively, for short baselines of a few kilometers, the survival probability can be parametrized as
\begin{equation}
\begin{aligned}
    P = 1-\cos^{4}\theta_{13}\sin^{2}2\theta_{12}\sin^{2}\Delta_{21} -\sin^{2}2\theta_{13}\sin^{2}\Delta_{ee}.
\end{aligned}
\end{equation}
Here, the effective mass-squared difference $\Delta m^{2}_{ee}$ is related to the wavelength of the oscillation observed at Daya Bay.

\begin{figure}
    \centering
    \includegraphics[width=0.5\linewidth]{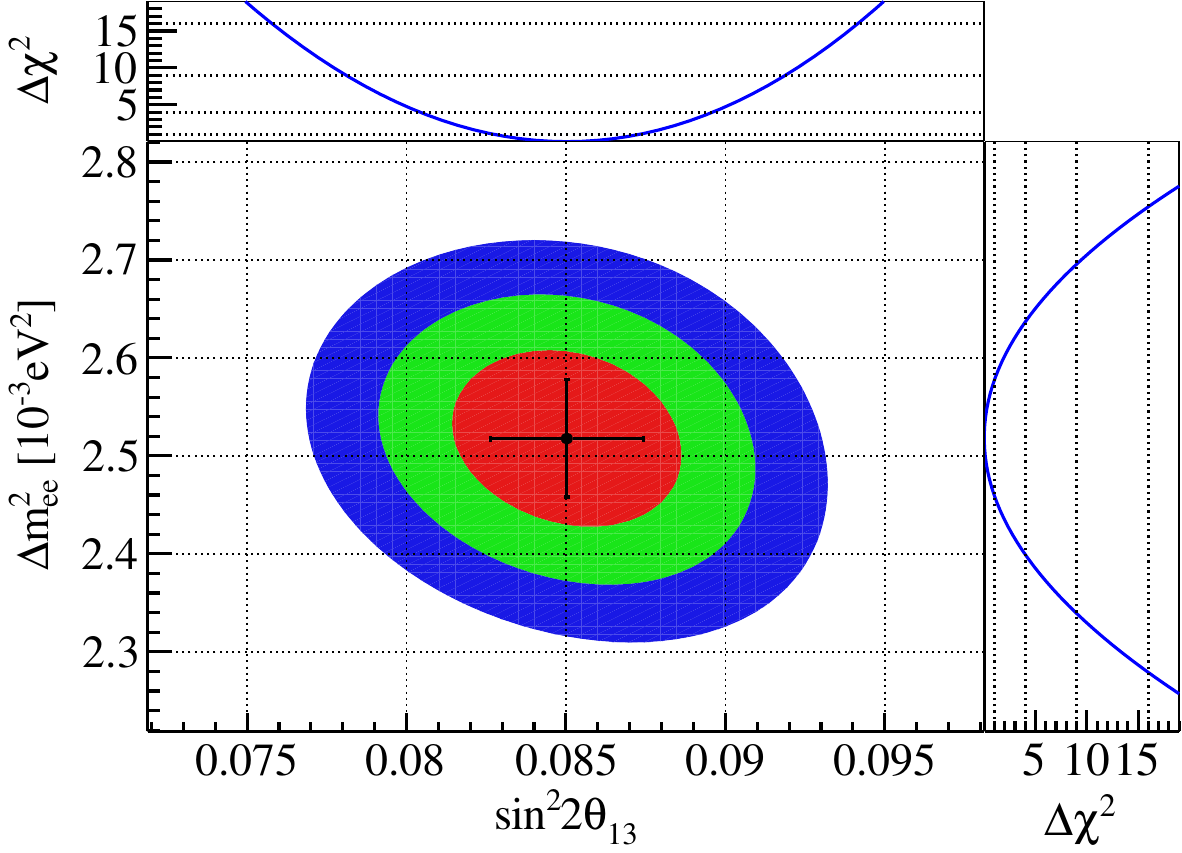}
    \caption{Error ellipses in the $\Delta m^{2}_{ee}-\sin^{2}2\theta_{13}$ space with the bestfit point indicated. The error bars display the one-dimensional 1-standard-deviation confidence intervals. The colored contours correspond to 1, 2, and 3 standard deviations.}
    \label{fig:contour}
\end{figure}

Fitting method B desribed in Ref. \cite{nGd2017PRD} was used in this work. 
Figure \ref{fig:contour} shows the covariance contours in the $\Delta m^{2}_{ee}-\sin^{2}2\theta_{13}$ space. The best-fit point with $\chi^{2}$/ndf = 559/517 yields $\sin^{2}2\theta_{13} = 0.0851 \pm 0.0024$, $\Delta m^{2}_{32} = (2.466 \pm 0.060) \times 10^{-3} \si{eV}^{2}$ for the normal mass ordering or $\Delta m^{2}_{32} = -(2.571 \pm 0.060) \times 10^{-3} \si{eV}^{2}$ for the inverted mass ordering \cite{nGd2023}.

\begin{figure}
    \centering
    \includegraphics[width=0.32\linewidth]{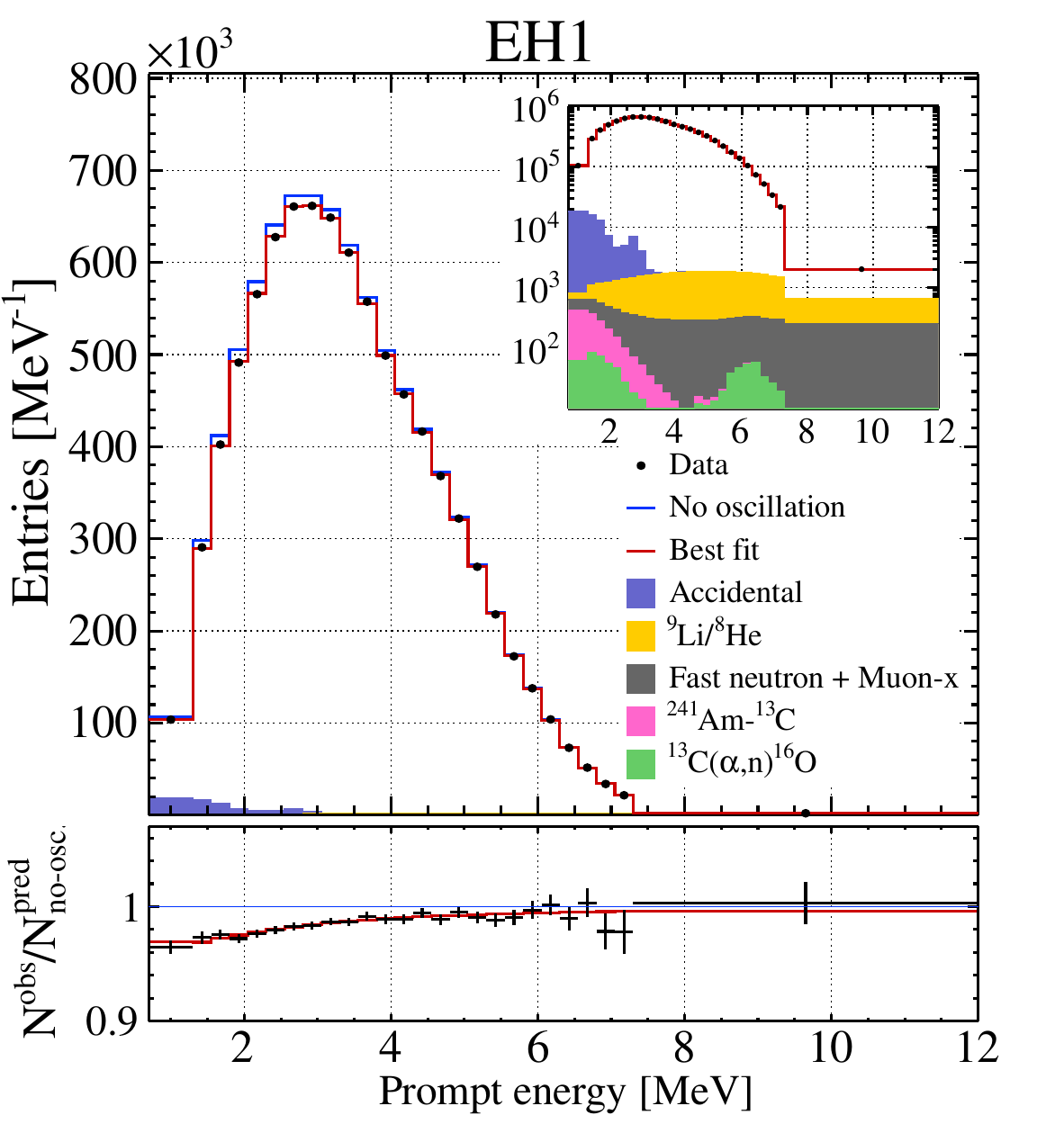}
    \includegraphics[width=0.32\linewidth]{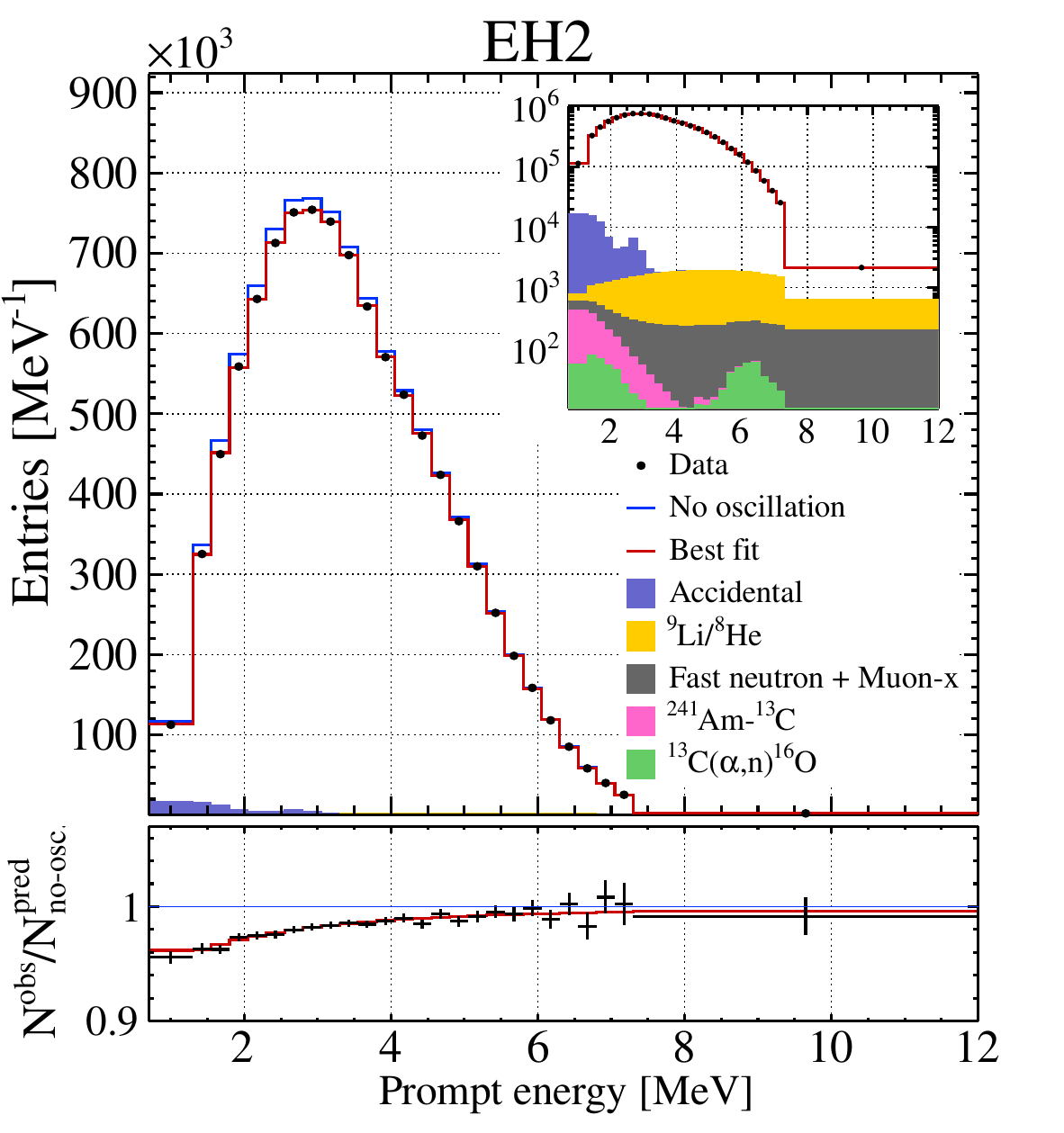}
    \includegraphics[width=0.32\linewidth]{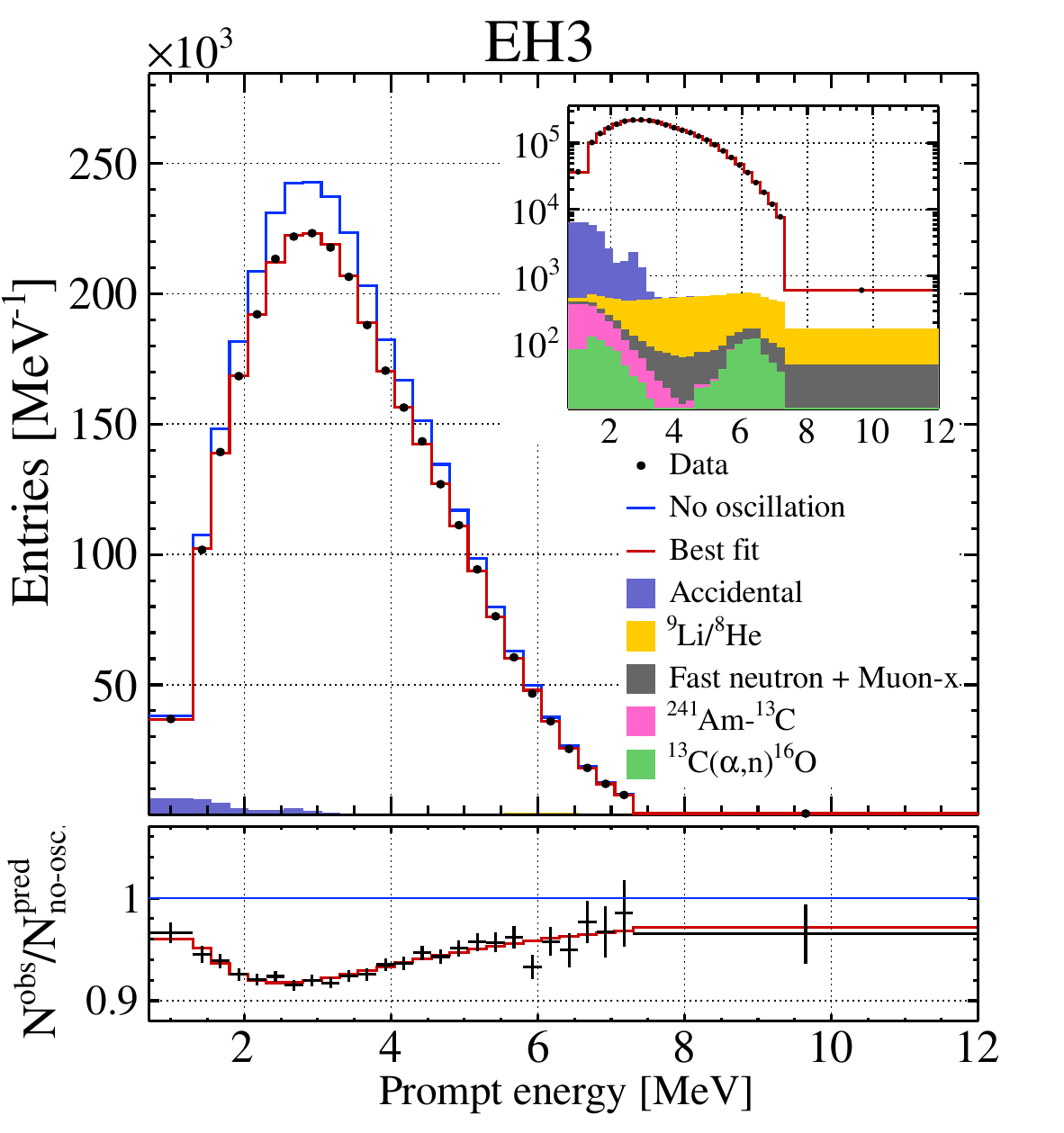}
    \caption{The measured prompt-energy spectra of EH1, EH2, and EH3 with the best-fit and no-oscillation curves superimposed in the upper panels. The shape of the backgrounds are apparent in the spectra with a logarithmic ordinate shown in the insets.}
    \label{fig:spectrum}
\end{figure}

As shown in Fig. \ref{fig:spectrum}, the best-fit prompt-energy distribution is in excellent agreement with the observed spectra in each experimental hall.

\begin{figure}
    \centering
    \includegraphics[width=0.6\linewidth]{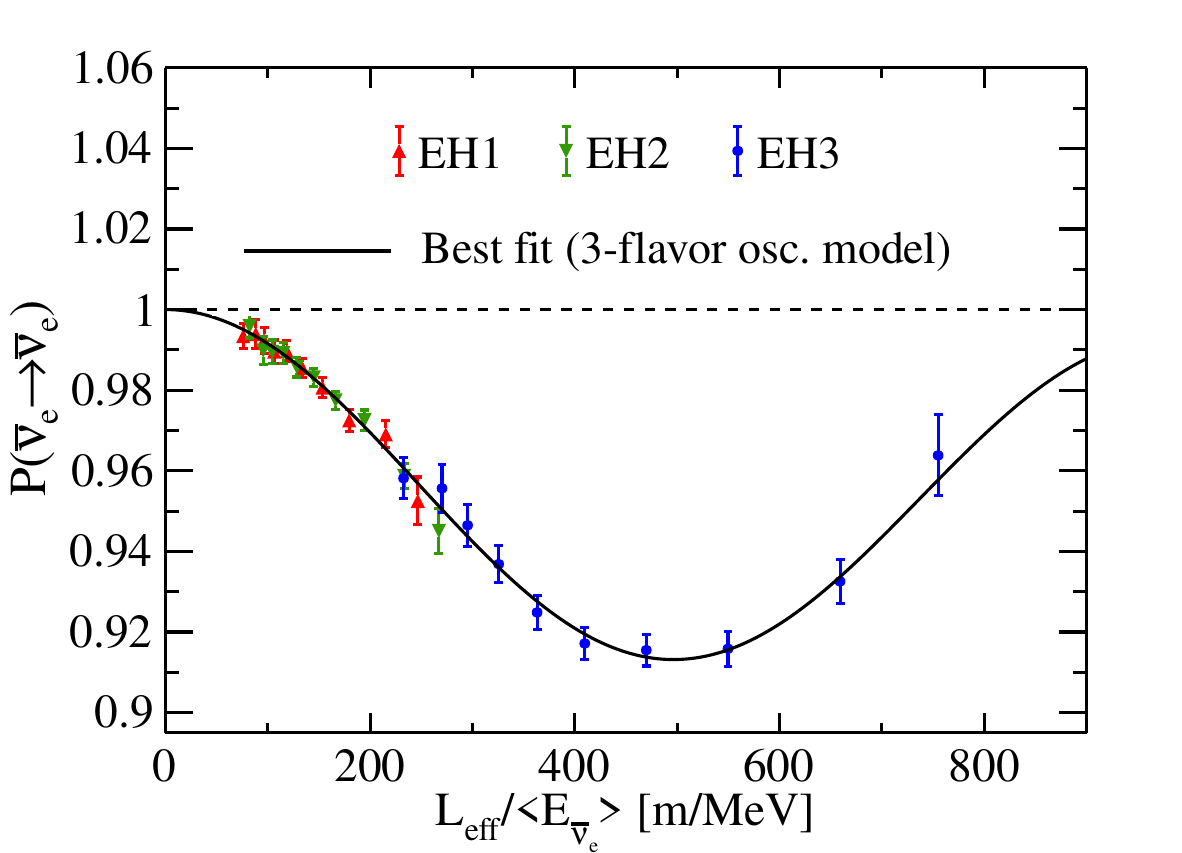}
    \caption{Measured disappearance probability as a function of the ratio of the effective baseline $L_{\rm eff}$ to the mean antineutrino energy $\langle E_{\overline{\nu}_{e}}\rangle$.}
    \label{fig:lovere}
\end{figure}

The normalized signal rate of the three halls as a function of $L_{\rm eff}/\langle E_{\overline{\nu}_{e}}\rangle$ with the best-fit curve superimposed is ploted in Fig. \ref{fig:lovere}, where $L_{\rm eff}$ and $\langle E_{\overline{\nu}_{e}}\rangle$ are the effective baseline and average $\overline{\nu}_{e}$ energy, respectively. The oscillation pattern related to $\theta_{13}$ is unambiguous.

\section{First Evidence of High Energy Reactor Antineutrinos}
High-energy reactor antineutrinos are likely generated by only a handful of short-lived $\beta$-decay nuclei with high end-point energies, such as $^{88,90}$Br and $^{94,96,98}$Rb. 
Nevertheless, for reactor antineutrino experiments, all previous measurements have focused on $E_p <$ 8 MeV, due to rare signals in the higher-energy region and large contamination with cosmogenic backgrounds. 
In spite of the rarity, high-energy reactor antineutrinos serve as a significant background for future measurements, such as the diffuse supernova neutrino background expected to permeate the Universe. Moreover, direct measurements of high-energy antineutrinos can provide a valuable new perspective for nuclear data validations relevant well beyond the bounds of neutrino physics.

In this section, we report the first measurement of high-energy reactor antineutrinos at Daya Bay, with nearly 9000 inverse beta decay candidates in the prompt energy region of 8–12 MeV from 1958 days of data collection \cite{heNu2022}.

\begin{figure}
    \centering
    \includegraphics[height=0.47\linewidth]{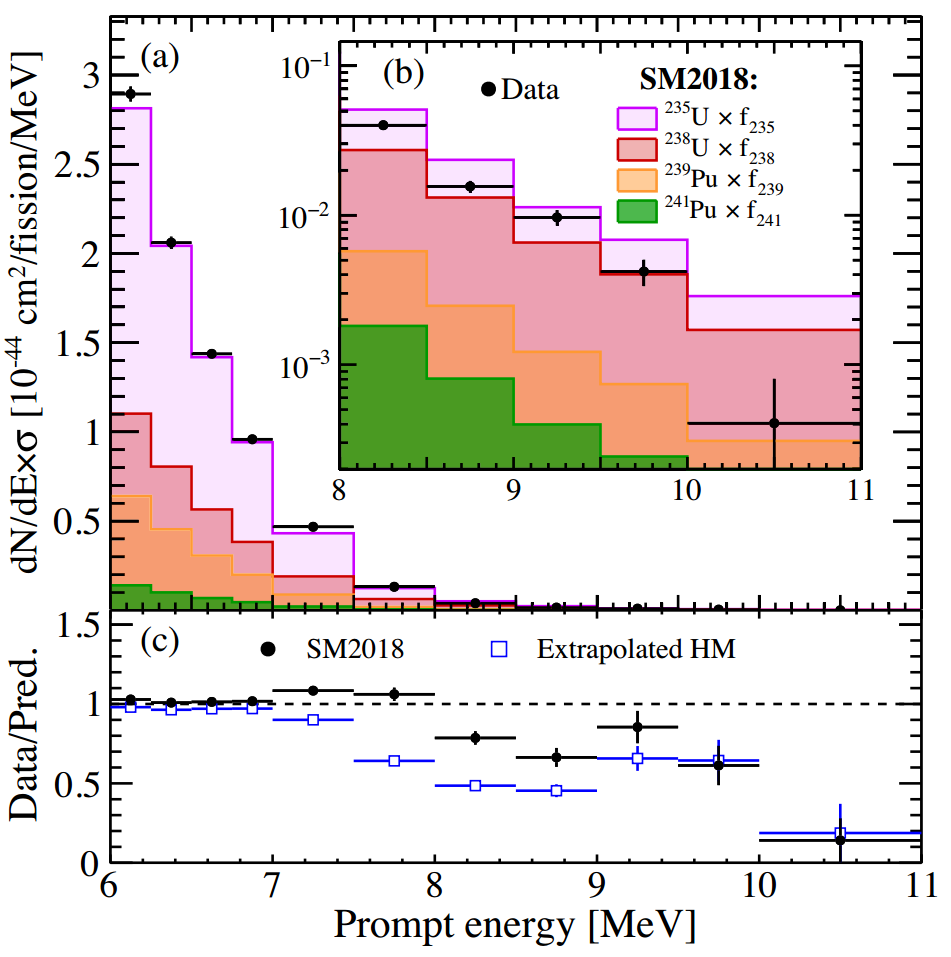}
    \includegraphics[height=0.47\linewidth]{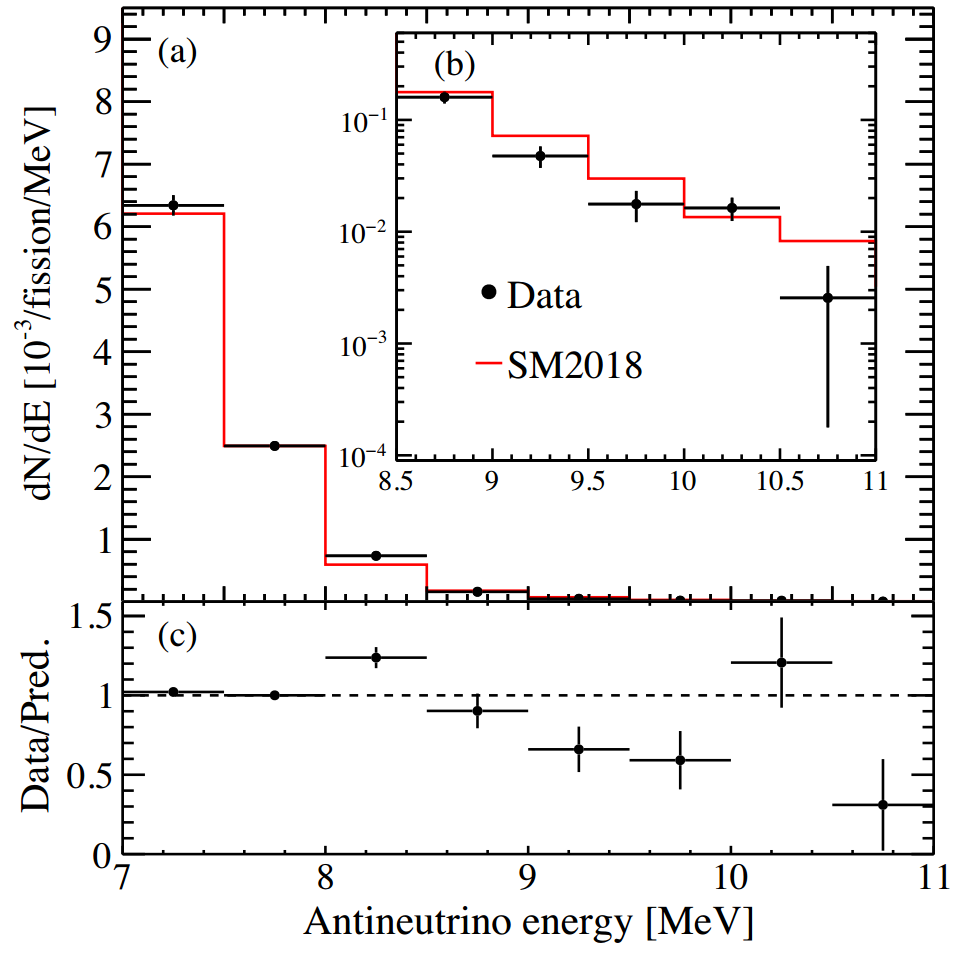}
    \caption{(Left) Measured prompt energy spectrum compared with the prediction from the SM2018 model. (Right) Reactor antineutrino energy spectrum above 7 MeV and its comparison with the prediction from SM2018.}
    \label{fig:heNuSpec}
\end{figure}

The main backgrounds with $E_p$ in 8–12 MeV are from muon decays, cosmogenic fast neutrons, and cosmogenic isotope decays. 
A multivariate analysis is performed to distinguish 2500 signal events from background statistically.
The hypothesis of no reactor antineutrinos with neutrino energy above 10 MeV is rejected with a significance of 6.2 standard deviations. 
We observed a 29\% antineutrino flux deficit in the prompt energy region of 8–11 MeV compared to a recent model prediction. 
Additionally, our work provide the unfolded antineutrino spectrum above 7 MeV as a data-based reference for other experiments, as shown in Fig. \ref{fig:heNuSpec}. 

\section{Summary}
To sum up, we have made a new measurement of $\sin^{2}2\theta_{13}$ with a precision of 2.8\% and the mass-squared differences reaching a precision of about 2.4\%. 
The reported $\sin^{2}2\theta_{13}$ will likely remain the most precise measurement of $\theta_{13}$ in the foreseeable future and be the crucial input for next-generation neutrino experiments studying the mass hierarchy and CP violation. 
Moreover, the hypothesis of no reactor antineutrinos with energy above 10 MeV is rejected with a significance of 6.2$\sigma$. 
More importantly, the energy region of reactor antineutrinos above 10 MeV is extended by direct measurement for the first time. 
Other latest results such as joint analysis of reactor antineutrino spectra with PROSPECT, and the measurement of the evolution of the flux and spectrum are not covered here, which can be seen in Ref. \cite{joint2022,evolution2023}. 


\end{document}